\documentclass[twocolumn,english,showpacs,floatfix,aps,pra]{revtex4}
\usepackage[latin1]{inputenc}
\usepackage{amsmath}
\usepackage{graphicx}
\usepackage{amssymb}

\usepackage{graphicx} 
\usepackage{dcolumn} 
\usepackage{bm} 
\usepackage{babel}

\newtheorem{definition}{Definition}

\begin{document}

\title{Operational Classification and Quantification of Multipartite
Entangled States}

\author{Gustavo Rigolin}
\email{rigolin@ifi.unicamp.br}

\author{Thiago R. de Oliveira}
\email{tro@ifi.unicamp.br}

\author{Marcos C. de Oliveira}
\email{marcos@ifi.unicamp.br}

\affiliation{Departamento de F\'{\i}sica da Mat\'{e}ria Condensada, Instituto
de F\'{\i}sica Gleb Wataghin, Universidade Estadual de Campinas,
Caixa Postal: 6165, cep 13083-970, Campinas, S\~{a}o Paulo, Brazil}

\begin{abstract}
We formalize and extend an operational
multipartite entanglement measure introduced in T. R. Oliveira, G.
Rigolin, and M. C. de Oliveira, Phys. Rev. A \textbf{73},
010305(R) (2006) through the generalization of global entanglement (GE) 
[ D. A. Meyer and N. R. Wallach, J. Math. Phys.
\textbf{43}, 4273 (2002)]. Contrarily to GE the
main feature of this new measure lies in the fact that we study
the mean linear entropy of \textit{all} possible partitions of a
multipartite system. This allows the construction of an operational 
multipartite entanglement measure which is able
to distinguish among different multipartite entangled states
that GE failed to discriminate. Furthermore, it is also
maximum at the critical point of the Ising chain in a transverse
magnetic field being thus able to detect a quantum phase transition.
\end{abstract}


\maketitle

\section{Introduction}

Since Schrödinger's seminal paper \cite{schroedinger}
entanglement is recognized to be at the heart of Quantum Mechanics
(QM). For a long time the study of entangled states was
restricted to the conceptual foundations of QM \cite{epr,bell}.
Since the last two decades, however, entanglement was also
recognized as a physical resource which can be used to
efficiently implement informational and computational tasks
\cite{livrodonielsen}. The understanding of the qualitative and
quantitative aspects of entanglement, therefore, naturally became
a fertile field of research.
Nowadays, entanglement of bipartite states (a joint state of a
quantum system partitioned in two subsystems $A$ and $B$) is quite
well understood. Good measures of entanglement for these systems
are available, specially for qubits \cite{wootters}. On the other hand
entanglement of multipartite states (a joint state of a quantum
system partitioned in more than two subsystems) cannot be understood
through simple extensions of the tools and measures employed for
bipartite entangled states. Most of the tools available to study
bipartite states (e.g. the Schmidt decomposition \cite{Schmidt})
are in general not useful for multipartite states. Even a
qualitative characterization of the many possible multipartite
entangled states (MES) is very complex since for a given
$N$-partitioned system there are many ``kinds'' of entanglement 
\cite{cirac,verschelde}. 
For example, let  $|\Psi\rangle_N = |\phi_{1}\rangle
\otimes \cdots \otimes |\phi_{p}\rangle \otimes
|\psi\rangle_{N-p}$ be a $N$-partite state in which
$|\phi_{i}\rangle$, $1\leq i \leq p$, is the $i$th subsystem state
and $|\psi\rangle_{N-p}$ is the state describing the other $N-p$
subsystems. If $|\psi\rangle_{N-p}$ is an entangled state then
$|\Psi\rangle$ is called a $p$-separable state \cite{footnote1}.
After discovering the value of $p$ for a given multipartite state
another complication shows up when we focus on
$|\psi\rangle_{N-p}$ since its subsystems can be entangled in
several inequivalent ways. For example, in the case of three
qubits there are two paradigmatic MES which cannot be converted to
each other via local operations and classical communication
(LOCC)\cite{cirac}. For four qubits, nine different kinds of
entanglement are possible, which cannot be converted to each other
via LOCC \cite{verschelde}. Thus after considerable work we still
lack a deep understanding of MES and new tools must be developed
in order to capture the essential features of genuine multipartite
entanglement (ME).

Our aim in this paper is to shed new light on the way 
ME is characterized and quantified. We intend to do this by
formalizing and extending an \textit{operational} ME measure
introduced in Ref. \cite{nossopaper}. We emphasize that it is an
operational measure in the sense that it is easily computable,
even for a multipartite state composed of many subsystems. This
new measure can be seen as an extension of the global entanglement
and we call it, from now on, the {\it generalized global
entanglement}: $E_{G}^{(n)}$. The generalized global entanglement
has several interesting features, two of which were already
explored in Ref. \cite{nossopaper}: (i) in contrast to the global
entanglement measure \cite{meyer} it can identify genuine MES
and (ii) it is maximum at the critical point for the Ising chain
in a transverse magnetic field.
Another important aspect of $E_{G}^{(n)}$ is the fact that it has
an intuitive physical interpretation. We can relate it to the
linear entropy of the pure state being studied as well as with the
purities of the reduced $n$-party states obtained by tracing out
the other $N-n$ subsystems \cite{viola,somma}.

This paper is organized as follows. In Sec.~\ref{definition} we
formally define
 $E_{G}^{(n)}$  and we extensively discuss a few important properties
satisfied by the generalized global entanglement. In
Sec.~\ref{examples} we calculate $E_{G}^{(n)}$ for the most
representatives MES. This gives us a good intuition of the meaning
of $E_{G}^{(n)}$ and illustrates its usefulness. We also compare
$E_{G}^{(n)}$ with other measures available, highlighting the main
differences and the advantages and disadvantages of each one. In
the same section we use $E_G^{(2)}$ to quantify the 
ground state multipartite entanglement of the one dimension (1D) Ising
model in a transverse magnetic field. Finally, in
Sec.~\ref{conclusion} we present our final remarks.

\section{Generalized Global Entanglement}
\label{definition}

Global entanglement (GE) was firstly introduced in Ref.
\cite{meyer} to quantify the ME contained in a chain of N qubits. Latter it
was demonstrated \cite{brennen} to be
equivalent to the mean linear entropy (LE) of all single qubits in the chain.
This connection between GE and LE considerably
simplified the calculation of GE and also extended it to systems
of higher dimensions.  An intuitive, though not so rigorous, way
of understanding GE is to consider it as quantifying the mean
entanglement between one subsystem with the rest of the
subsystems. In this process we are dividing a system of $N$
components into a single subsystem and the remaining $N-1$
subsystems. We could, nevertheless, separate the system into two
partition blocks, one containing $L$ subsystems and the other one
$N-L$ \cite{latorre,latorre2}. There are many different ways to
construct a given ``block''. In Refs. \cite{latorre,latorre2} a
block of $L$ subsystems consisted of the first $L$ successive
subsystems: $L=\{ S_1, S_2, S_3, \ldots, S_L\}$. But any other
possible combination of $L$ subsystems could be employed to
construct a block. We may have, for instance, a block formed by
the first $L$ odd subsystems: $L=\{S_1, S_3, S_5, \ldots,
S_{2L-1}\}$. It is legitimate to compute the LE of each one of
these possible partitions. Roughly speaking this allows us to
detect and quantify all possible `types' of entanglement in a
multipartite pure state. The generalized global entanglement
$(E_{G}^{(n)})$ is defined to take into account all of those
possible partitions of a system composed of $N$ subsystems. Before
we define $E_{G}^{(n)}$ we highlight two of its main important
qualities: (a) It is a relatively simple and operational measure.
Since it is based on LE it can be easily evaluated and it is valid
for any type of multipartite pure state (states belonging either
to finite or infinite dimension Hilbert spaces); (b) Each class of
$E_{G}^{(n)}$ is related to the mixedness/purity of \textit{all}
possible $n$-partite reduced density matrices out of a system
composed of $N$ subsystems, and thus it is not restricted to
reduced density matrices of only one subsystem as the original GE
\cite{meyer,somma,viola}. This fact is helpful for the physical
understanding of $E_{G}^{(n)}$.

Following the definition of $E_{G}^{(n)}$ we move to the study of
the general properties of this new measure relating it to the
mixedness/purity of the various reduced density matrices of the
system.  After that we particularize to qubits focusing on the
ability of the generalized global entanglement to classify and
quantify MES. We conclude this section by presenting a variety of
examples, which clarify the necessity of all the classes of
$E_{G}^{(n)}$, i. e. $n=1,2,3,\ldots$, to properly understand the
many facets of MES.

\subsection{Formal Definition of the Measure}

Consider a system $S$ which is partitioned into $N$ subsystems
$S_{i}$, $1 \leq i \leq N$. Let $|\Psi\rangle \in \mathcal{H}$ be
a quantum state describing $S$ and $\mathcal H$ the Hilbert space
of the whole system. Since we have $N$ subsystems, $\mathcal{H} =
\mathcal{H}_{1} \otimes \cdots \otimes \mathcal{H}_{N} =
\bigotimes_{i=1}^{N}\mathcal{H}_{i}$, in which $\mathcal{H}_{i}$
is the Hilbert space associated with $S_{i}$. The density matrix
of $S$ is $\rho=|\Psi\rangle \langle \Psi|$ and we define the
generalized global entanglement \cite{CommentScott} as,
\begin{eqnarray}
E_{G}^{(n)}(\rho) & = & \frac{1}{C^{N-1}_{n-1}}\sum_{i_1=1}^{N-1}
\sum_{i_2=i_1 + 1}^{N-1}
\sum_{i_3=i_2 + 1}^{N-1} \cdots \nonumber \\
& & \cdots \sum_{i_{n-1}=i_{n-2}+1}^{N-1}  G(n,i_{1},i_{2},
\ldots, i_{n-1}), \label{egn}
\end{eqnarray}
where all the parameters are natural numbers, $n < N$, and
\begin{displaymath}
C^{N-1}_{n-1} =\frac{(N-1)!}{(N-n)!(n-1)!}
\end{displaymath}
is the definition of the binomial coefficient. Note that the
summation is over all $i_{k}$'s, with the restriction that $1\leq
i_{1}<i_{2}<\cdots<i_{n-1} \leq N - 1$. We also assume $i_0=0$.
The function $G$ is given as,
\begin{widetext}
\begin{eqnarray}
G(n,i_{1},i_{2}, \ldots, i_{n-1})  =  \frac{d}{d-1}\left[ 1 -
\frac{1}{N-i_{n-1}}  \sum_{j=1}^{N-i_{n-1}}
\text{Tr}\left( \rho^{2}_{j,j+i_{1},j+i_{2}, \ldots, j + i_{n-1}} \right)
\right],
\label{gn}
\end{eqnarray}
\end{widetext}
where $\rho_{j,j+i_{1},j+i_{2}, \ldots, j + i_{n-1}}$ is obtained
by tracing out all the subsystems but $S_A = \{S_{j}, S_{j+i_{1}},
S_{j+i_{2}}, \ldots, S_{j+i_{n-1}}\}$ and $d =
\text{min}\{\text{dim}\:S_A, \text{dim}\:\overline{S}_A \}$. Here
$\text{dim}\:S_{A}$ and $\text{dim}\:\overline{S}_{A}$ are,
respectively,
 the Hilbert space dimension of the subsystem $S_{A}$ and of its complement
$\overline{S}_A$. In resume the index $n$ is for the number of
subsystems in the $A$ partition and the indexes
$i_1,i_2,...,i_{n-1}$ are the neighborhood addressing for each of
the involved subsystems.

\subsection{General Properties}

Eqs.~(\ref{egn}) and (\ref{gn}) are valid for any multipartite
pure system, even systems described by continuous variables
(Gaussian states for example). The key concept behind generalized
global entanglement is the fact that it is based on the linear
entropy, which is an entanglement monotone easily calculated for
the vast majority of pure states. Thus, by its very definition,
$E_{G}^{(n)}$ and $G$ inherit all the properties satisfied by LE,
including the crux of all entanglement monotones: non-increase
under LOCC.

Another important concept of $E_{G}^{(n)}$ and $G$ is the
introduction of classes of multipartite entanglement (ME) labeled
by the index $n$. As we will see, they are all related with the
many ways a multipartite state can be entangled. Moreover, a
genuine $n$-partite entangled state must have non-zero
$E_{G}^{(n)}$ and $G$'s for all classes $n$. Here a \emph{genuine}
MES means a multipartite pure entangled system in which no pure
state can be defined to anyone of its subsystems. There is only
one pure state describing the whole joint system. For three
qubits, for instance, the states $|GHZ\rangle=(1/\sqrt{2})
(|000\rangle + |111\rangle)$ and
$|W\rangle=(1/\sqrt{3})(|001\rangle + |010\rangle + |100\rangle)$
are genuine MES but $|\xi\rangle = (1/\sqrt{2})(|00\rangle +
|11\rangle)|0\rangle$ is not.

Let us now explicitly show how the first classes of $E_{G}^{(n)}$ look like.
This will clarify the physical meaning of the measure as well as the
intuitive aspects
which led us to arrive at the general and formal definitions given in
Eqs.~(\ref{egn}) and (\ref{gn}).

\subsubsection{First Class: $n = 1$}

When $n=1$ Eqs. (\ref{egn}) and (\ref{gn}) are the same,
\begin{equation}
E_{G}^{(1)}(\rho) = G(1) =   \frac{d}{d-1}\left[ 1 -
\frac{1}{N}  \sum_{j=1}^{N}
\text{Tr}\left( \rho^{2}_{j} \right)
\right],
\label{eg1}
\end{equation}
and if we remember the definition of the linear entropy for the
subsystem $j$ \cite{brennen,indianos},
\begin{equation}
E_{L}(\rho_{j})=\frac{d}{d-1}\left[1-\text{Tr}\left(\rho_{j}^{2}\right)\right],
\label{linear1}
\end{equation}
then Eq.~(\ref{eg1}) can be written as \cite{nossopaper}
\begin{equation}
E_{G}^{(1)}=\frac{1}{N}\sum_{j=1}^{N}E_{L}(\rho_{j})=
\langle E_{L}(\rho_j)\rangle.
\label{mean1}
\end{equation}
In other words, $E_{G}^{(1)}$ is simply the mean linear entropy of
all the subsystems $S_j$. We should mention that for qubits
($d=2$), $E_{G}^{(1)}$ was shown \cite{brennen} to be  exactly the
Meyer and Wallach global entanglement \cite{meyer}.

The physical intuition behind the study of the mean linear
entropies lies in the fact that the more a state is a genuine MES
the more mixed their reduced density matrices should be. However,
we should not limit ourselves to evaluating the reduced density
matrices of single subsystems $S_{j}$. We can take either two, or three, ...,
 or $n$ subsystems and calculate their reduced density matrices and
also calculate their mean linear entropies. This is the reason of
why we introduced the other classes of generalized global
entanglement.

\subsubsection{Second Class: $n = 2$}

For $n=2$ Eqs.~(\ref{egn}) and (\ref{gn}) are not identical anymore, being,
nevertheless, entanglement monotones:
\begin{eqnarray}
E_{G}^{(2)}(\rho) & = & \frac{1}{N-1}\sum_{i_1=1}^{N-1} G(2,i_{1}),
\label{eg2} \\
G(2,i_{1})  & = & \frac{d}{d-1}\left[ 1 -
\frac{1}{N-i_{1}}  \sum_{j=1}^{N-i_{1}}
\text{Tr}\left( \rho^{2}_{j,j+i_{1}} \right)
\right]. \label{g2}
\end{eqnarray}
Now we deal with the reduced joint density matrix for subsystems $S_j$
and $S_{j+i_{1}}$. The extra parameter $i_{1}$ is introduced to
take account of the many possible `distances' between the two
subsystems. For nearest neighbors $i_1=1$, next-nearest neighbors
$i_2=2$, and so forth.

Noticing that the linear entropy of the subsystems $S_{j}$ and
$S_{j+i_{1}}$ by tracing out the rest of the other subsystems is
given by
\begin{equation}
E_{L}(\rho_{j,j+i_{1}})=\frac{d}{d-1}\left[1-\text{Tr}
\left(\rho_{j,j+i_{1}}^{2}\right)\right],
\label{linear2}
\end{equation}
then Eq.~(\ref{g2}) can be written as,
\begin{equation}
  G(2,i_1)=
\frac{1}{N-i_1}\sum_{j=1}^{N-i_1} E_{L}(\rho_{j,j+i_1}) =\langle
E_{L}(\rho_{j,j+i_1})\rangle. \label{media2}
\end{equation}
This implies that Eq.~(\ref{eg2}) is simply given by
\begin{equation}
  E_{G}^{(2)}(\rho)=\frac{1}{N-1}\sum_{i_1=1}^{N-1}
\langle E_{L}(\rho_{j,j+i_1})\rangle= \langle\langle
E_{L}(\rho_{j,j+i_1})\rangle\rangle, \label{media_da_media2}
\end{equation}
where the double brackets represent the averaging over all possible  
$G(2,i_1)$, $1\leq i_1\leq N-1$.

Looking at Eqs.~(\ref{media2}) and (\ref{media_da_media2}) we can
easily interpret $E_{G}^{(2)}$ and $G(2,i_{1})$. First, let us
deal with $G(2,i_{1})$. We assume that all the subsystems are organized
in a linear chain. (This assumption simplifies the discussion in
what follows.) If we remember that $1\leq
i_{1}\leq N-1$, where $N$ is the number of subsystems,
Eq.~(\ref{media2}) tells us that $G(2,i_{1})$ is nothing but the
mean linear entropy of two subsystems with the rest of the other
subsystems conditioned on that these two subsystems are $i_{1}$
lattice sites apart.

\begin{figure}[!ht]
\includegraphics[width=2.25in]{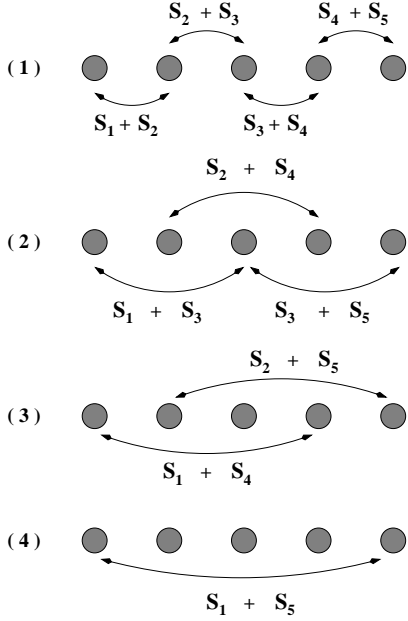}
\caption{\label{g2_5qubits_fig} All combinations of two elements
out of five.}
\end{figure}

For concreteness, let us explicitly write all the possible
$G(2,i_{1})$ for a linear chain of five subsystems. Since $N=5$
we have $1\leq i_{1}\leq 4$, which gives four $G$'s pictorially
represented in Fig.~\ref{g2_5qubits_fig}:
\begin{itemize}
\item[(1)] $G(2,1)$, which is the mean linear entropy (LE) of
the following pairs of subsystems with the rest of the chain:
$\{(S_1,S_2),(S_2,S_3),(S_3,S_4),(S_4,S_5)\}$;
\item[(2)]$G(2,2)$, which is the mean LE of the
following pairs of subsystems:
$\{(S_1,S_3),(S_2,S_4),(S_3,S_5)\}$;

\item[(3)]$G(2,3)$, which is the mean LE of the
following pairs of subsystems:
$\{(S_1,S_4),(S_2,S_5)\}$;

\item[(4)]$G(2,4)$, which is the mean LE of the
following pairs of subsystems:
$\{(S_1,S_5)\}$.
\end{itemize}

Finally, Eq.~(\ref{media_da_media2}) shows that $E_{G}^{(2)}$ is
the mean linear entropy of two subsystems with the rest of the
chain irrespective of the distance between the two subsystems,
{\it i.e.}, it is the averaged summation of all the (1)-(4) kinds
of $G(2,i_1)$, $1\le i_1\le 4$.

\subsubsection{Third Class: $n=3$}

By setting $n=3$ Eqs.~(\ref{egn}) and (\ref{gn}) become
\begin{equation}
E_{G}^{(3)}(\rho) = \frac{2}{(N-1)(N-2)}\sum_{i_1=1}^{N-1}
\sum_{i_2=i_1 + 1}^{N-1} G(n,i_{1},i_{2}), \label{eg3}
\end{equation}
and
\begin{equation}
G(3,i_{1},i_{2})  =  \frac{d}{d-1}\left[ 1 - \frac{1}{N-i_{2}}
\sum_{j=1}^{N-i_{2}} \text{Tr}\left( \rho^{2}_{j,j+i_{1},j+i_{2}}
\right) \right]. \label{g3}
\end{equation}
Eq.~(\ref{g3}) deals with reduced density matrices of three
subsystems: $S_{j}$, $S_{j+i_1}$, and $S_{j+i_2}$. Therefore,
$G(3,i_{1},i_{2})$ is the mean linear entropy of all three
subsystems with the rest of the chain conditioned to that
$S_{j+i_1}$ and $S_{j+i_2}$ are, respectively, $i_1$ and $i_2$
lattice sites apart from $S_{j}$. Taking the mean of all possible
$G(3,i_{1},i_{2})$ we obtain Eq.~(\ref{eg3}). This is equivalent
to averaging over all linear entropies of three subsystems
irrespective of their distances. Although we do not explicitly write them
here, similar expressions as those given by Eqs.~(\ref{media2})
and (\ref{media_da_media2}) can be obtained for this class.

\begin{figure}[!ht]
\includegraphics[width=3.2in]{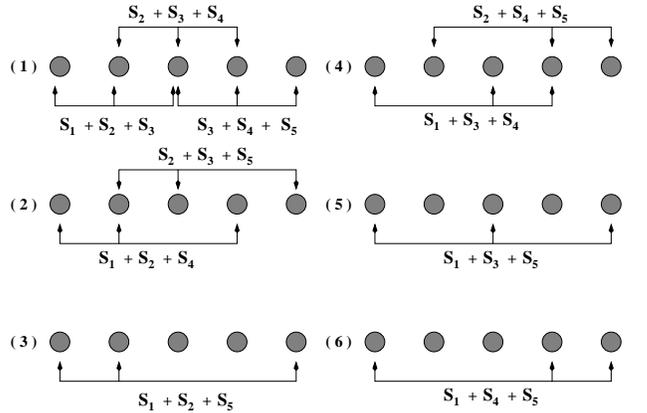}
\caption{\label{g3_5qubits_fig} All combinations of three elements
out of five.}
\end{figure}

Again, as we did for the second class, it is explanatory to
analyze in details the $N=5$ case. Now $1\leq i_{1}< i_{2}\leq 4$.
This time we have six $G$'s (See Fig.~\ref{g3_5qubits_fig}):
\begin{itemize}
\item[(1)] $G(3,1,2)$, which is the mean linear entropy (LE) of
the following triples of subsystems with the rest of the chain:
$\{(S_1,S_2,S_3),(S_2,S_3,S_4),(S_3,S_4,S_5)\}$;
\item[(2)] $G(3,1,3)$, which is the mean LE of the following
triples of subsystems: $\{(S_1,S_2,S_4),(S_2,S_3,S_5)\}$;

\item[(3)] $G(3,1,4)$, which is the mean LE of the following
triples of subsystems: $\{(S_1,S_2,S_5)\}$;

\item[(4)] $G(3,2,3)$, which is the mean LE of the following
triples of subsystems: $\{(S_1,S_3,S_4), (S_2,S_4,S_5)\}$;

\item[(5)] $G(3,2,4)$, which is the mean LE of the following
triples of subsystems: $\{(S_1,S_3,S_5)\}$;

\item[(6)] $G(3,3,4)$, which is the mean LE of the following
triples of subsystems: $\{(S_1,S_4,S_5)\}$.
\end{itemize}

\subsubsection{Higher Classes: $n \geq 4$}

Remembering that $n<N$, higher classes $n$ of $E_G^{(n)}(\rho)$
only make sense for systems such that $N\ge n+1$ subsystems. The
higher a class $n$ the greater the number of $G$'s necessary for
the computation of $E_G^{(n)}(\rho)$. This is a satisfactory
property we should expect from a useful multipartite entanglement
measure since as we increase the number of partitions of a system
we increase the way it may be entangled \cite{cirac,verschelde}.

If we employ the definition of LE for $n$ subsystems out of a
total of $N$,
\begin{equation}
E_{L}(\rho_{j,\ldots,j+i_{n-1}})=\frac{d}{d-1}\left[1-\text{Tr}
\left(\rho_{j,\ldots,j+i_{n-1}}^{2}\right)\right], \label{linearn}
\end{equation}
we can write Eqs.~(\ref{gn}) and (\ref{egn}) respectively as
\begin{eqnarray}
G(n,i_1, \ldots, i_{n-1})&=& \langle E_{L}(\rho_{j,j+i_1, \ldots,
i_{n-1}})\rangle, \label{median} \\
E_{G}^{(n)}(\rho)&=& \langle\langle E_{L}(\rho_{j,j+i_1, \ldots,
i_{n-1}}) \rangle\rangle. \label{media_da_median}
\end{eqnarray}
In Eq.~(\ref{median}) the single pair of brackets $\langle \;
\rangle$ represents the averaging over all possible configurations
of $n$ subsystems in which subsystem $S_{j+i_{k}}$ is $i_{k}$
lattice sites apart from $S_j$. Here $1<k<n-1$. Finally, the
double brackets $\langle\langle \; \rangle\rangle$ is the average
of the linear entropy of $n$ subsystems over all possible
combinations (distances) in which they can be arranged.

We should mention at this point that $E_{G}^{(n)}$ and $G$ are
more general than the block entanglement ($E_{B}^{(n)}$) as
presented in Refs.~\cite{scott,latorre,latorre2}. By block
entanglement it is understood that we divide a set of $N$
subsystems $\{ S_{1}, S_{2}, \ldots, S_{N}\}$ in two blocks,
$A_{n} = \{ S_{1}, S_{2}, \ldots, S_{n}\}$ and $B_{N-n}= \{ S_{n+1},
S_{n+2}, \ldots, S_{N}\}$, and calculate the linear or von Neumann
entropy between blocks $A_{n}$ and $B_{N-n}$. In the language of
generalized global entanglement, block entanglement for a
translational symmetric state is simply
\begin{displaymath}
E_{B}^{(n)} = G(n,i_1=1,i_2=1,\ldots,i_{n-1}=1),
\end{displaymath}
which is only one of the many $G$'s we can define. The main
difference between these two measures lies in the fact that we
allow all possible combinations of $n$ subsystems out of $N$ to
represent a possible `block'. Contrarily to block entanglement,
here there exists no restriction onto the subsystems belonging to
a given `block' to be nearest neighbors. They lie anywhere in the
system's domain.

\subsection{Particular Properties for Qubits}

Although Eqs.~(\ref{egn}) and (\ref{gn}) are defined for Hilbert
spaces of arbitrary dimensions we now focus on some properties of
$E_{G}^{(n)}$ and $G$ for qubits. There are two main reasons for
studying qubits in detail. Firstly, they are recognized as a key
concept for quantum information theory and secondly, the simplest
multipartite states are constructed employing qubits.

Let $\rho = |\Psi \rangle \langle \Psi |$ be the density matrix of
a $N$ qubit system and $\rho_{j}=\text{Tr}_{\overline{j}}(\rho)$
the reduced density matrix of subsystem $S_{j}$, which is obtained
by tracing out all subsystems but $S_j$. A general one qubit
density matrix can be written as
\begin{equation}
\rho_{j}=\text{Tr}_{\overline{j}}(\rho)= \frac{1}{2}\sum_{\alpha}
p^{\alpha}_{j}\sigma_{j}^{\alpha}, \label{um_qubit}
\end{equation}
where the coefficients are given by
\begin{equation}
p^{\alpha}_{j}=\text{Tr}\left(\sigma_{j}^{\alpha}\rho_{j}\right)=
\langle\Psi|\sigma_{j}^{\alpha}|\Psi\rangle. \label{coeficiente1}
\end{equation}
Here $\sigma_{j}^{\alpha}$ is the Pauli matrix acting on the site
$j$, $\alpha = 0, x, y, z$, where $\sigma^{0}$ is the identity
matrix of dimension two, and $p^{\alpha}_{j}$ is real. Since $\rho_j$
is normalized $p_0 = 1$.
Using Eqs.~(\ref{um_qubit}) and (\ref{coeficiente1}) we obtain
\begin{equation}
\text{Tr}\left(\rho_j^2 \right) = \frac{1}{2} \left( 1 + \langle
\sigma^{x}_j \rangle^2 +\langle \sigma^{y}_j \rangle^2 +  \langle
\sigma^{z}_j \rangle^2 \right). \label{traco1}
\end{equation}
This last result implies that Eq.~(\ref{eg1}) can be written as
\begin{equation}
E_{G}^{(1)} = 1 - \frac{1}{N}\sum_{j=1}^{N} \left( \langle
\sigma^{x}_j \rangle^2 + \langle \sigma^{y}_j \rangle^2 + \langle
\sigma^{z}_j \rangle^2  \right). \label{ent_correlacao1}
\end{equation}

One interesting situation occurs when we have translational
invariant states $\rho$. (The Ising model ground state for
example.) In this scenario $\langle \sigma_{i}^{\alpha} \rangle =
\langle \sigma_{j}^{\alpha}\rangle$ for any $i$ and $j$.
Therefore, Eq.~(\ref{ent_correlacao1}) becomes
\begin{equation}
E_{G}^{(1)} = 1 - \langle \sigma^{x}_j \rangle^2 - \langle
\sigma^{y}_j \rangle^2 - \langle \sigma^{z}_j \rangle^2,
\label{translation-g1}
\end{equation}
which is related to the total magnetization $M$ of the system,
$|M|^2= N(\langle \sigma_{j}^x \rangle^2+\langle
\sigma_{j}^y\rangle^2+ \langle \sigma_{j}^z\rangle^2)$, by
$E_{G}^{(1)} = 1 - \frac{|M|^2}{N}$.

By tracing out all subsystems but $S_{i}$ and $S_{j}$ we obtain
the two qubit reduced density matrix
\begin{equation}
\rho_{ij}=\text{Tr}_{\overline{ij}}(\rho)=
\frac{1}{4}\sum_{\alpha,\beta}
p^{\alpha\beta}_{ij}\sigma_{i}^{\alpha}\otimes\sigma_{j}^{\beta},
\label{reduzida2}
\end{equation}
where
\begin{equation}
p^{\alpha\beta}_{ij}=\text{Tr}\left(\sigma_{i}^{\alpha}\sigma_{j}^{\beta}
\rho_{ij}\right)= \langle
\Psi|\sigma_{i}^{\alpha}\sigma_{j}^{\beta}|\Psi\rangle.
\label{coeficientes2}
\end{equation}
Eq.~(\ref{reduzida2}) is the most general way to represent a
two-qubit state and together with Eq.~(\ref{coeficientes2}) imply
that
\begin{equation}
\text{Tr}\left( \rho_{ij}^2 \right) = \frac{1}{4}
\sum_{\alpha,\beta}\langle \sigma_{i}^{\alpha} \sigma_{j}^{\beta}
\rangle^{2}. \label{traco2}
\end{equation}
Remark that in Eq.~(\ref{traco2}) the trace of $\rho_{ij}^{2}$ is
the sum of all one and two-point correlation functions. Moreover,
since $E_{G}^{2}$ and $G(2,i_{1})$ depend on Eq.~(\ref{traco2}),
we find in these entanglement measures both diagonal and
off-diagonal correlation functions.

Again it is instructive to study translational symmetric states in
which $p_{ij}^{\alpha\beta}=p_{ij}^{\beta\alpha}$ for any $\alpha$
and $\beta$. Using
this assumption in Eq.~(\ref{g2}) we get
\begin{eqnarray}
G(2,i_{1}) & = & 1 - \frac{2}{3}\left[
\langle\sigma_{j}^{x}\rangle^2 + \langle\sigma_{j}^{y}\rangle^2 +
\langle\sigma_{j}^{z}\rangle^2
 + \langle\sigma_{j}^{x}\sigma_{j+i_{1}}^{y}\rangle^2
\right. \nonumber \\
& & + \langle\sigma_{j}^{x}\sigma_{j+i_{1}}^{z}\rangle^2 +
\langle\sigma_{j}^{y}\sigma_{j+i_{1}}^{z}\rangle^2
+ \langle\sigma_{j}^{x}\sigma_{j+i_{1}}^{x}\rangle^2/2   \nonumber \\
& & \left. + \langle\sigma_{j}^{y}\sigma_{j+i_{1}}^{y}\rangle^2/2
+ \langle\sigma_{j}^{z}\sigma_{j+i_{1}}^{z}\rangle^2/2 \right].
\label{translation-g2}
\end{eqnarray}
Note that the previous formula is not valid for $N\leq 3$. For
$N=2$ only $E_{G}^{(1)}$ is defined and for $N=3$ we have $d =
\text{min}\{\text{dim}\:S_A, \text{dim}\:\overline{S}_A \}=2$ and
not $d=4$, the value of $d$ for all $N\geq 4$. Now if we compare
$G(2,i_{1})$ with the concurrence (a bipartite entanglement
monotone), as we do for the Ising model in Sec. \ref{infinite-chains},
we will note that while the
concurrence does not depend on any one-point and on any off-diagonal
two-point correlation function \cite{nome_dificil1,nome_dificil2}
$G(2,i_{1})$ does.

\subsection{Why Do We Need Higher Classes?}
\label{HigherClasses}

The simple fact that different types of entanglement appear as we
increase the number of qubits (or equivalently the number of
subsystems) \cite{cirac,verschelde} indicates that the various
classes here introduced may be useful to classify and quantify the
many facets of ME. For example, the first class $E_G^{(1)}$
does not suffice to unequivocally quantify MES. Although it is
maximal for Greenberger-Horne-Zeilinger (GHZ) states \cite{ghz} it
is also maximal for a state which is not a MES, as we now
demonstrate. Let us compute $E_{G}^{(1)}$ for three paradigmatic
multipartite states. The first one is the GHZ state:
\begin{equation}
|GHZ_{N}\rangle=\frac{1}{\sqrt{2}} \left(|0\rangle^{\otimes
N}+|1\rangle^{\otimes N}\right),
\end{equation}
 where $|0\rangle^{\otimes N}$ and $|1\rangle^{\otimes N}$ represent,
respectively, $N$ tensor products of the states $|0\rangle$ and
$|1\rangle$. The GHZ state is a genuine MES since by measuring
only one of the qubits in the standard basis we know exactly the
results of the other $N-1$ qubits. Furthermore, tracing out any
one of the qubits we obtain a separable state. A direct
calculation gives $E_{G}^{(1)}(GHZ_{N})=1$.

The second state we shall analyze is given by a tensor product of
$N/2$ Einstein-Podolsky-Rosen (EPR) Bell states \cite{viola}:
\begin{equation}
|EPR_{N}\rangle=|\Phi^{+}\rangle\otimes\cdots\otimes|\Phi^{+}\rangle
=|\Phi^{+}\rangle^{\otimes\frac{N}{2}},
\end{equation}
 where $|\Phi^{+}\rangle=(1/\sqrt{2})(|00\rangle+|11\rangle)$. For
definiteness, we chose one specific Bell state. However, the
results here derived are quite general and valid for any $N/2$
tensor products of Bell states. This state is obviously not a
genuine MES. Only the pairs of qubits $(2j-1,2j)$, where $j=1,2,...,N$,
are entangled. Nevertheless, we again obtain
$E_{G}^{(1)}(EPR_{N})=1$. This last result illustrates that $E_{G}^{(1)}$ 
being maximal is not a sufficient condition to detect
genuine MES. Note that $E_{G}^{(1)}$ for both the $GHZ_{N}$ and
$EPR_{N}$ states are independent of the number of qubits $N$ in
the chain.

The last state we consider is the W state \cite{cirac}. It is
defined as,
\begin{equation}
|W_{N}\rangle=\frac{1}{\sqrt{N}}\sum_{j=1}^{N}|000\cdots1_{j}\cdots000\rangle.
\end{equation}
 The state $|000\cdots1_{j}\cdots000\rangle$ represents a $N$ qubit
state in which the $j$-th qubit is $|1\rangle$ and all the others
are $|0\rangle$. As shown in Ref. \cite{meyer},
$E_{G}^{(1)}(W_{N})=4(N-1)/N^{2}$. Note that $E_{G}^{(1)}$ depends
on $N$ and at the thermodynamic limit ($N\rightarrow\infty$) we
have $E_{G}^{(1)}(W_{N})=0$. For three qubits, the W state was
shown \cite{cirac} to be a genuine MES not convertible via LOCC to
a GHZ state.

\begin{figure}[!ht]
\includegraphics[width=2.5in]{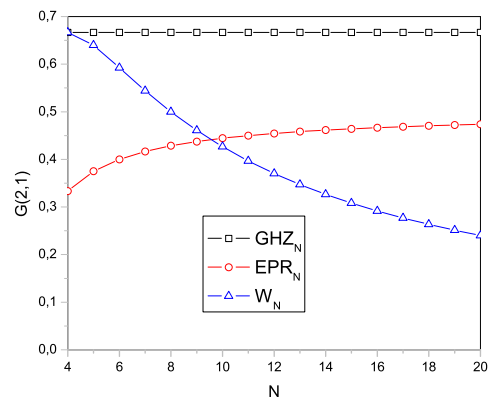}
\caption{\label{figuraG21} (Color online) Here we show $G(2,1)$ as
a function of the number of qubits $N$ for the states $GHZ_{N}$,
$EPR_{N}$ and $W_{N}$. Note that only when $N=4$ we have two
states with the same entanglement. Furthermore, for $4\leq
N\leq8$, $W_{N}$ is more entangled than $EPR_{N}$. This ordering
is changed for $N\geq9$.}
\end{figure}

\begin{figure}[!ht]
\includegraphics[width=2.5in]{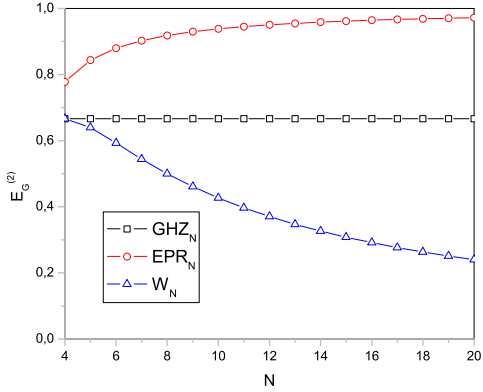}
\caption{\label{figuraGE2} (Color online) Here we show
$E_{G}^{(2)}$ as a function of $N$. Again, only when $N=4$ we have
two states with the same entanglement. Moreover, for $N\geq4$,
$EPR_{N}$ is the most entangled state.}
\end{figure}

The computation of $E_{G}^{(2)}$ and $G(2,1)$ give different
values for each of those states. Remark that for $N=2$ the
previous functions are not defined and that for $N=3$ $E_{G}^{(2)}
= G(2,1) = 1$. Table \ref{tabela1} shows $E_{G}^{(2)}$ and
$G(2,1)$ for the states $GHZ_{N},EPR_{N}$, and $W_{N}$. We should
mention that due to translational symmetry, $G(2,1)$ and
$E_{G}^{(2)}$ are identical for the states $GHZ_{N}$ and $W_{N}$.
\begin{table}[!ht]
\caption{\label{tabela1} The third and fourth columns give
$G(2,1)$ and $E_{G}^{(2)}$ for the three states listed in the
first column when $N > 3$. The second column gives $E_{G}^{(1)}$
for all $N$. Contrary to $E_{G}^{(1)}$, we see that $G(2,1)$ and
$E_{G}^{(2)}$ distinguish the three states from each other.}
\begin{ruledtabular}
\begin{tabular}{cccc}

 & $E_{G}^{(1)}$ & $G(2,1)$ & $E_{G}^{(2)}$ \\ \hline

 & & & \\

 $GHZ_{N}$ & $1$ & \large{$\frac{2}{3}$} & \large{$\frac{2}{3}$} \\

 & & & \\

 $EPR_{N}$ & $1$ &\large{$\frac{N-2}{2(N-1)}$} &
\large{$\frac{(2N-1)(N-2)}{2(N-1)^{2}}$} \\

& & & \\

 $W_{N}$ & \large{$\frac{4(N-1)}{N^{2}}$} &
\large{$\frac{16(N-2)}{3N^{2}}$} &
\large{$\frac{16(N-2)}{3N^{2}}$}
\end{tabular}
\end{ruledtabular}
\end{table}
It is interesting to note that depending on the value of $N$, the
states are differently classified through $G(2,1)$. Fig.
\ref{figuraG21} illustrates the behavior of $G(2,1)$ for those
three paradigmatic state as we vary $N$.
A similar behavior is observed for $E_{G}^{(2)}$ (Fig.
\ref{figuraGE2}). In this case, however, $EPR_{N}$ is the most
entangled state for long chains. The reason for this lies in the
definition of $E_{G}^{(2)}$. For the $EPR_{N}$ state, $G(2,l)$ = 1
for any $l\geq2$. Therefore, since $E_{G}^{(2)}$ is obtained
averaging over all $G(2,l)$, for long chains $G(2,1)$ does not
contribute much and $E_{G}^{(2)}\rightarrow1$.

We also calculated the values of $E_{G}^{(1)}$, $E_{G}^{(2)}$, and
$G(2,1)$ at the thermodynamic limit. See Tab. \ref{tabela2}.
\begin{table}[!ht]
\caption{\label{tabela2} $E_{G}^{(1)}$, $G(2,1)$, and
$E_{G}^{(2)}$ at the thermodynamic limit.}
\begin{ruledtabular}
\begin{tabular}{cccc}
$N \rightarrow \infty$ & $E_{G}^{(1)}$ & $G(2,1)$ & $E_{G}^{(2)}$
\\ \hline  $GHZ_{N}$ & $1$ & $2/3$ & $2/3$ \\  $EPR_{N}$ & $1$ &
$1/2$ & $1$ \\  $W_{N}$ & $0$ & $0$ & $0$
\end{tabular}
\end{ruledtabular}
\end{table}
Thus even at the thermodynamic limit $E_{G}^{(2)}$ and $G(2,1)$
distinguish the three states. The ordering of the states,
nevertheless, is different. Again this is related to the
definition of $E_{G}^{(2)}$ and is due to the contribution of
$G(2,l)$, $l\geq2$, in the calculation of $E_{G}^{(2)}(EPR_{N})$.

\begin{figure}[!ht]
\includegraphics[width=3.0in]{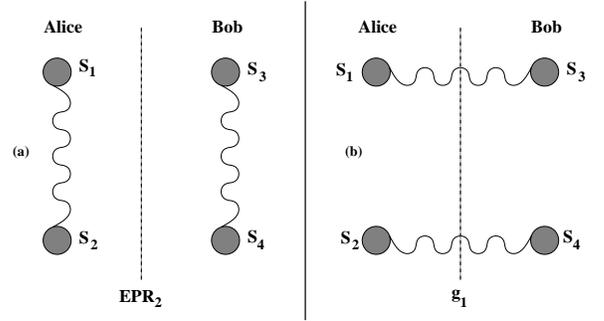}
\caption{\label{g1_and_epr2} Pictorial representations of the
states (a) $EPR_2$ and (b) $g_1$.}
\end{figure}

\begin{figure}[!ht]
\includegraphics[width=3.1in]{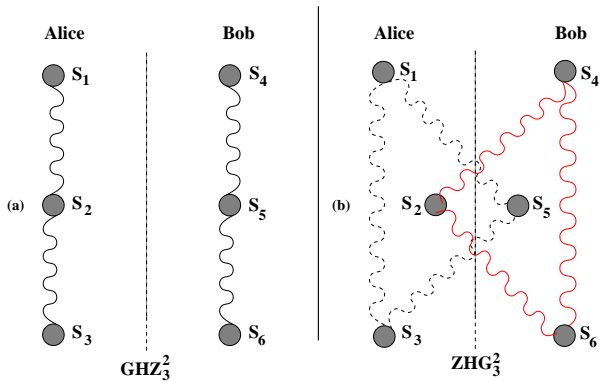}
\caption{\label{ghz_and_zhg} (Color online) Pictorial
representations of the states (a) $GHZ^{2}_{3}$ and (b)
$ZHG^{2}_{3}$.}
\end{figure}

Besides a measure of multipartite entanglement being able to
distinguish different kinds of states it should not differentiate
states that essentially contain the same amount of entanglement.
For example, let us consider the following state,
\begin{eqnarray}
|EPR_2\rangle &=& |\Phi^+\rangle_{12}|\Phi^+\rangle_{34} \nonumber \\
&=& \frac{1}{2}(|0\rangle_1|0\rangle_2|0\rangle_3|0\rangle_4 +
|0\rangle_1|0\rangle_2|1\rangle_3|1\rangle_4 \nonumber \\
&&+ |1\rangle_1|1\rangle_2|0\rangle_3|0\rangle_4 +
|1\rangle_1|1\rangle_2|1\rangle_3|1\rangle_4).
\end{eqnarray}
This state describes a pair of EPR states where subsystem $S_1$ is
entangled with $S_2$ and $S_3$ is entangled with $S_4$. Consider
now the state
 defined as \cite{rigolintele}
\begin{eqnarray}
|g_1\rangle &=&
\frac{1}{2}(|0\rangle_1|0\rangle_2|0\rangle_3|0\rangle_4 +
|0\rangle_1|1\rangle_2|0\rangle_3|1\rangle_4 \nonumber \\
&&+|1\rangle_1|0\rangle_2|1\rangle_3|0\rangle_4 +
|1\rangle_1|1\rangle_2|1\rangle_3|1\rangle_4 ) \nonumber \\
& = &  \frac{1}{2}(|0\rangle_1|0\rangle_3|0\rangle_2|0\rangle_4 +
|0\rangle_1|0\rangle_3|1\rangle_2|1\rangle_4 \nonumber \\
&&+|1\rangle_1|1\rangle_3|0\rangle_2|0\rangle_4 +
|1\rangle_1|1\rangle_3|1\rangle_2|1\rangle_4 ) \nonumber \\
&=&|\Phi^+\rangle_{13}|\Phi^+\rangle_{24},
\end{eqnarray}
which is also a pair of EPR states. This time, however, subsystem $S_1$ is
entangled with $S_3$ and subsystem $S_2$ is entangled with $S_4$
(See Fig. \ref{g1_and_epr2}).
Although different pairs of subsystems are entangled in these two
different states, their amount of entanglement is the same: there
are two EPR states in both cases. This fact is captured by the
entanglement measures here introduced, {\it i.e.}
$E_{G}^{(n)}(EPR_2)=E_{G}^{(n)}(g_1)$. The block entanglement,
nevertheless, does not always give the same value for the two
states above (see Tab. \ref{tabela_block}).
\begin{table}[!ht]
\caption{\label{tabela_block} Comparison between $E_{G}^{(n)}$,
$G(2,1)$, and $E_{B}^{(n)}$}
\begin{ruledtabular}
\begin{tabular}{cccccc}
 & $E_{G}^{(1)}$ & $E_{G}^{(2)}$ & $G(2,1)$ & $E_{B}^{(1)}$ &
$E_{B}^{(2)}$ \\ \hline
$EPR_{2}$ & $1$ & $7/9$ & $1/3$  & $1$ & $0$ \\
$g_1$ & $1$ & $7/9$ & $1/3$  & $1$ & $1$
\end{tabular}
\end{ruledtabular}
\end{table}
This example illustrates that the block entanglement, as its name
suggests, quantifies only the entanglement of partition $A$ (sites
$1$ and $2$) with partition $B$ (sites $3$ and $4$). The
generalized global entanglement $E_{G}^{(n)}$, however, quantifies
the amount of entanglement of a state independently on the way it
is distributed among the subsystems. We can go further and show
the importance of using higher classes $E_G^{(n)}$ to correctly
quantify the entanglement of a multipartite state no matter how
the entanglement is distributed among the subsystems. For example,
consider the state
\begin{equation}
|GHZ_{N}^{M}\rangle = |GHZ_{N}\rangle^{\otimes M},
\end{equation}
where the integer $M\geq 1$ represents how many tensor products of
$GHZ_{N}$ we have. Restricting ourselves to $N=3$ and $M=2$ we
get,
\begin{eqnarray}
|GHZ_{3}^{2}\rangle &=& \frac{1}{\sqrt{2}} \left(|000\rangle +
|111\rangle \right) \otimes  \frac{1}{\sqrt{2}}
\left(|000\rangle + |111\rangle \right) \nonumber \\
&=& \frac{1}{2}( |000000\rangle + |000111\rangle
+ |111000\rangle \nonumber \\
& & +  |111111\rangle ). \label{ghz3}
\end{eqnarray}
Here, subsystems $S_1$, $S_2$, and $S_3$ form a genuine MES and
$S_4$, $S_5$, and $S_6$ another one. For this state
$E_{B}^{(3)}(GHZ^{2}_{3})=0$. If we interchange the second qubit
($S_2$) with the fifth one ($S_5$) we obtain the following state:
\begin{eqnarray}
|ZHG_{3}^{2}\rangle &=& \frac{1}{2}( |000000\rangle +
|010101\rangle
+ |101010\rangle \nonumber \\
& & +  |111111\rangle ).
\end{eqnarray}
Now subsystems $S_1$, $S_3$, and $S_5$ form a genuine MES and
$S_2$, $S_4$, and $S_6$ another one (See Fig. \ref{ghz_and_zhg}).
Those two states have the same amount of entanglement, {\it i. e.}
two GHZ states. However, the computation of the block entanglement
gives $E_B^{(3)}(ZHG_{3}^{2})=6/7\neq E_B^{(3)} (GHZ^{2}_{3})$.
Had we employed the generalized global entanglement we would have
obtained $E_{G}^{(3)}(GHZ^{2}_{3})=
E_{G}^{(3)}(ZHG_{3}^{2})$ instead. In general we have
$E_{B}^{(n)}(GHZ^{2}_{n})\neq E_{B}^{(n)}(ZHG_{n}^{2})$ and
$E_{G}^{(n)}(GHZ^{2}_{n})=E_{G}^{(n)}(ZHG_{n}^{2})$. Therefore, if
we want to study the amount of entanglement of a multipartite
state, independently on how it is distributed among the
subsystems, we should employ $E_{G}^{(n)}$ instead of
$E_{B}^{(n)}$, since the later furnishes only the amount of
entanglement between a particular two block-partition in which
the system can be divided.

\section{Usefulness of the Generalized Global Entanglement}
\label{examples}

In this section we present two examples in which we explore the
ability of $E_{G}^{(n)}$ and the auxiliary measure $G(n, i_1, i_2,
\ldots, i_{n-1})$ to quantify multipartite entanglement. The first
example deals with a finite chain of four qubits. We show that
$E_{G}^{(2)}$ together with $G(2,1)$ allow us to correctly
identify MES. Moreover, comparing the values of $G(2,i_1)$ for all
the MES here presented we are led to a practical definition of
what is a genuine MES. In the second example we investigate the
entanglement properties of the Ising model ground state. We show
that $E_{G}^{(2)}$ and $G(2,i_{1})$ are maximal at the critical
point and we analyze what correlation functions are responsible
for this behavior of the generalized global entanglement. The
results herein presented suggest that the long range correlations
in the critical point for the Ising model are related to genuine
MES.

\subsection{Finite Chains}
\label{finite}

Let us now focus on the simplest non-trivial spin-1/2 chain, {\it
i. e.} states with $N=4$ qubits, by studying the entanglement
properties of four genuine MES \cite{osterloh,chua}. The first one
\cite{osterloh} is the famous four qubit GHZ state \cite{ghz},
\begin{equation}
|GHZ_{4}\rangle = |\Phi_{1}\rangle = \frac{1}{\sqrt{2}}\left(
|0000\rangle + |1111\rangle \right).
\end{equation}
Qualitative and quantitative features of this state were already
discussed in Sec. \ref{HigherClasses}. A direct calculation gives
$ E_{G}^{(1)}\left(\Phi_{1}\right) = 1;
E_{G}^{(2)}\left(\Phi_{1}\right) =
G(2,i_{1})\left(\Phi_{1}\right) = 2/3, $
where $i_{1}=1,2,3$. The second state \cite{osterloh} is written
as,
\begin{eqnarray}
|\Phi_{2}\rangle &=& \frac{1}{\sqrt{6}}\left( \sqrt{2}|1111\rangle
+
|1000\rangle + |0100\rangle +|0010\rangle \right. \nonumber \\
& & \left. + |0001\rangle \right).
\end{eqnarray}
Calculating its first and second order generalized global
entanglement we obtain
$ E_{G}^{(1)}\left(\Phi_{2}\right) = 1;
E_{G}^{(2)}\left(\Phi_{2}\right) = G(2,i_{1})\left(\Phi_{2}\right)
= 8/9. $ Note that as well as $|\Phi_{1}\rangle$ this state is a
translational symmetric state. Moreover, $G(2,i_{1})(\Phi_{2})\geq
G(2,i_{1})(\Phi_{1})$. This last result will turn out to be very
useful in constructing an operational definition of MES. The third
state \cite{osterloh} is given as,
\begin{equation}
|\Phi_{3}\rangle = \frac{1}{2}\left(|1111\rangle + |1100\rangle +
|0010\rangle +|0001\rangle \right).
\end{equation}
Since this state is not translational symmetric, $G(2,i_{1})$ are
not all equal. After a straightforward calculation we obtain
$ E_{G}^{(1)}\left(\Phi_{3}\right) = 1;
E_{G}^{(2)}\left(\Phi_{3}\right) =  25/27;
G(2,1)\left(\Phi_{3}\right) = 7/9; G(2,2)\left(\Phi_{3}\right) =
G(2,3)\left(\Phi_{3}\right) = 1. $
Again we should note that $G(2,i_{1})(\Phi_{3})\geq
G(2,i_{1})(\Phi_{1})$.

These three states have in common a few remarkable properties
\cite{osterloh}: (a) The local density operator describing each
qubit is the maximally mixed state $(1/2)I_{2}$, where $I_{2}$ is
the $2\times2$ identity matrix, thus explaining  why
$E_{G}^{(1)}=1$ for all of them. (b) The two- and three-qubits
reduced operators do not have any $k$-tangle \cite{coffman},
$k=2,3$. This emphasizes that they all are genuine MES, i. e.
there is no pairwise or triplewise entanglement. (c) They cannot
be transformed into one another by LOCC.

We shall consider a fourth state,
\begin{eqnarray}
|\chi\rangle &=& \frac{1}{2\sqrt{2}} \left( |0000\rangle -
|0011\rangle
- |0101\rangle + |0110\rangle \right. \nonumber \\
& & \left. + |1001\rangle + |1010\rangle + |1100\rangle +
|1111\rangle \right),
\end{eqnarray}\label{st42}
recently introduced and extensively studied in Ref. \cite{chua}.
The main feature of this state lies in its usefulness to teleport
an arbitrary two-qubit state. Employing $\chi$ this task can be
accomplished either from subsystems $S_1$ and $S_{2}$ to $S_{3}$
and $S_{4}$ or from $S_{1}$ and $S_{3}$ to $S_{2}$ and $S_{4}$.
The usual channel (two Bell states) used to teleport an arbitrary
two-qubit state \cite{rigolintele,guo} can teleport two qubits
only from a specific location to another one: from  $S_1$ and
$S_{2}$ to $S_{3}$ and $S_{4}$ for example. In addition state
$|\chi\rangle$ has a hybrid behavior in the sense that it resembles
both the $GHZ$ and $W$ states \cite{chua}. Tracing out any one of
the qubits the remaining reduced density matrix $\sigma$ has
maximal entropy, a characteristic of the $GHZ$ state. However,
$\sigma$ has a non-zero negativity \cite{negativity} between one
qubit and the other two \cite{chua}, a property of the $W$ state.
By calculating the generalized global entanglement we obtain
$ E_{G}^{(1)}\left(\chi\right) = 1; E_{G}^{(2)}\left(\chi\right) =
23/27; G(2,1)\left(\chi\right) = 8/9; G(2,2)\left(\chi\right) = 1;
G(2,3)\left(\chi\right) = 2/3. $ Again we see that for
\textit{all} $i_{1}$ we have $G(2,i_{1})(\chi) \geq
G(2,i_{1})(\Phi_{1})$.

We have grouped in Tab. \ref{tabela4qubits} the entanglement
calculated for the previous four states.
\begin{table}[!ht]
\caption{\label{tabela4qubits} Calculated values of $E_{G}^{(n)}$
and $G(2,i_1)$ for the genuine MES shown in Sec. \ref{finite} and
for the $EPR_{2}$ state.}
\begin{ruledtabular}
\begin{tabular}{cccccc}
 & $E_{G}^{(1)}$ & $E_{G}^{(2)}$ & $G(2,1)$ & $G(2,2)$ &
$G(2,3)$ \\ \hline
$EPR_{2}$ & $1$ & $7/9 \approx 0.778$ & $1/3$  & $1$ & $1$ \\
$\Phi_{1}$ & $1$ & $2/3 \approx 0.667$ & $2/3$  & $2/3$ & $2/3$ \\
$\Phi_{2}$ & $1$ & $8/9 \approx 0.889$ & $8/9$  & $8/9$ & $8/9$ \\
$\Phi_{3}$ & $1$ & $25/27 \approx 0.926$ & $7/9$ & $1$ & $1$ \\
$\chi$ & $1$ & $23/27 \approx 0.852$ & $8/9$  & $1$ & $2/3$
\end{tabular}
\end{ruledtabular}
\end{table}
It is clear then that $E_G^{(1)}$ cannot be considered as the last
word concerning the quantification and classification of MES. A
glimpse of the first column in Tab. \ref{tabela4qubits} shows that
all the five states listed have $E_{G}^{(1)}=1$, even the
$EPR_{2}$ state, an obvious non-genuine MES. Therefore, since
$E_{G}^{(1)}=1$ is not useful to classify different genuine MES or
to correctly identify them we are compelled to go further and
study the higher classes of the generalized global entanglement in
order to achieve such a goal. Turning our attention to
$E_{G}^{(2)}$ we see that it is different for \textit{all} the
five states listed in Tab. \ref{tabela4qubits}, implying that
$E_{G}^{(2)}$ can distinguish among the five states. According to
$E_{G}^{(2)}$ the most entangled state is $\Phi_{3}$, which was
shown to be a genuine MES \cite{osterloh}.

Moreover, important clues for the understanding of what kind of
entanglement is present in a given multipartite state are also
available in $G(2,i_{1})$, $i_{1}=1,2,3$. Actually, these
auxiliary  entanglement measures give us a more detailed view of
the types of entanglement a state has than $E_{G}^{(2)}$ since the
latter is an average over all $G(2,i_{1})$. For example, if we
relied only on $E_{G}^{(2)}$ to decide whether or not a state is a
genuine MES we would arrive at a wrong answer. This point is
clearly demonstrated if we compare $E_{G}^{(2)}$ for the states
$EPR_2$ and $\Phi_1$ ($GHZ_4$). Looking at Tab.
\ref{tabela4qubits} we see that $E_{G}^{(2)}(EPR_2)>
E_{G}^{(2)}(\Phi_1)$, where $EPR_2$ is not a genuine MES. The
averaging process, as explained in Sec. \ref{HigherClasses}, is
responsible for this relatively high value of $E_{G}^{(2)}$ for
the state $EPR_2$. Remark that for translational symmetric states
$E_{G}^{(2)}$ and $G(2,i_1)$ are equivalent to detect a genuine
MES. However, if we analyze all the $G(2,i_{1})$ terms we are able
to detect a common characteristic shared only by the genuine MES:
for $\chi$ and all $i_1$ we have $G(2,i_1)(\Phi_j,\chi)\geq G(2,i_1)(GHZ_4) =
2/3$. This suggests the following operational definition of a
genuine MES:
\begin{definition}
Let $|\Psi\rangle$ be a pure state describing four qubits. If
$G(1)=1$ and $G(2,i_1)(\Psi)\geq G(2,i_1)(GHZ_4)=2/3$,
$i_1=1,2,3$, then $|\Psi\rangle$ is a genuine MES. \label{def1}
\end{definition}

Besides being practical, Definition \ref{def1} has a simple
physical interpretation if we remember that $E_{G}^{(2)}$ and
$G(2,i_1)$ are constructed in terms of the linear entropy of any
two qubits with the rest of the chain. Noticing that the linear
entropy is related to the purities of the two-qubit reduced
density matrices, the definition above establishes an upper bound
for all the two-qubit purities of a MES. In other words, if all
the two-qubit purities are below this upper bound the $N$ qubit
state can be considered a genuine MES \cite{footnote2}.
Furthermore, this upper bound was chosen to be that of the $GHZ$
state, which is undoubtedly a genuine MES.

Remark also that since $G(2,i_1)$ is a monotonically decreasing
function of the purities, an upper bound for the purities implies
a lower bound for the value of $G(2,i_1)$ (cf.
Definition \ref{def1}). We can easily generalize this definition
to $N$ qubits if we express it in terms of all $n$-qubit purities
($n<N$):
\begin{definition}
A pure state of N qubits $|\Psi\rangle$ is a genuine MES if
\begin{eqnarray*}
\text{Tr}\left(\rho^2_{j_1}\right) &\leq& \text{Tr}\left(
\sigma^2_{1}
\right)=1/2, \\
\text{Tr}\left(\rho^2_{j_1,j_2}\right) &\leq& \text{Tr}\left(
\sigma^2_{1,2}
\right)=1/2, \\
 & \vdots &  \\
\text{Tr}\left(\rho^2_{j_1,j_2,\ldots, j_n}\right) &\leq&
\text{Tr}\left( \sigma^2_{1,2, \ldots, n}\right)=1/2,
\end{eqnarray*}
where
\begin{eqnarray*}
\rho_{j_1, j_2, \ldots, j_n}&=&\text{Tr}_{\overline{j_1, j_2
\ldots,
j_n}}\left(|\Psi\rangle\langle\Psi|\right),\\
\sigma_{1,2, \ldots, n}&=&\text{Tr}_{\overline{1,2, \ldots, n}}
\left(|GHZ_N\rangle\langle GHZ_N|\right),
\end{eqnarray*}
and
\begin{eqnarray*}
1\leq j_1 \leq N,\\
1\leq j_1 < j_2 \leq N,
\end{eqnarray*}
\begin{eqnarray*}
\vdots
\end{eqnarray*}
\begin{eqnarray*}
1\leq j_1 < j_2 < \cdots < j_{n}\leq N.
\end{eqnarray*}
\label{def2}
\end{definition}
Note that as we increase the size of the chain we need to
calculate more and more purities. Take for instance the state
$GHZ^2_3$ given by Eq.~(\ref{ghz3}). A direct calculation gives
$\text{Tr}(\rho^2_{j_1})=1/2$ for $1\leq j_1\leq 6$,
$\text{Tr}(\rho^2_{3,4})=1/4$, and
$\text{Tr}(\rho^2_{j_1,j_2})=1/2$ for all $1 \leq j_1<j_2\leq 6$
but $(j_1,j_2)=(3,4)$. Hence, if we restricted Definition
\ref{def2} just to the one- and two-qubits reduced density matrices
we would erroneously conclude that $GHZ^2_3$ is a genuine MES.
Extending, however, the definition to all possible reduced density
matrices we can detect that $GHZ^2_3$ is not a genuine MES since
$\text{Tr}(\rho^2_{1,2,3})=1$, a clear violation of Definition
\ref{def2}.

We end this section remarking that Definition \ref{def2} is
completely defined only for finite chains. For infinite chains
($N\rightarrow \infty$) one would have to calculate all $G(n,i_1,
i_2)$ (and $G(n, i_1, i_2, \ldots, i_{n-1})$) to completely
characterize a genuine $n$-partite entangled state. Finally, the
previous definition does not imply that all genuine MES must have
$\text{Tr}\left(\rho^2_{j}\right)\leq 1/2$,
$\text{Tr}\left(\rho^2_{j,j+i_1}\right)\leq 1/2$, $\ldots$,
$\text{Tr}\left(\rho^2_{j,j+i_1, \ldots, j+i_{n-1}}\right)\leq
1/2$. It is thus only a sufficient condition for a state to be a
genuine MES.

\subsection{Infinite Chains}
\label{infinite-chains}

Currently there is an increasing interest on the relation between
entanglement and Quantum Phase Transitions occurring in infinite
spin chains \cite{latorre,latorre2,nielsen,Nature,tognetti,verstraete,venuti}).  For spin
chains presenting a second order quantum phase transition (QPT) the
correlation length goes to infinity at the critical point, thus
suggesting interesting entanglement properties for the ground state of such
models.
Particularly interesting is the 1D Ising model \cite{Ising
original}, which is translationally invariant and presents a
ferromagnetic-paramagnetic QPT. As we have seen in Sec. II, the
generalized global entanglement is easily evaluated for a system
with translational symmetry. In this perspective, for the 1D Ising
model ground state, here we compute $G(1)$, which is shown to
behave similarly to the von Neumann entropy calculated in Ref.
\cite{nielsen}, and $G(2,i_1)$ for some values of $i_1$.

The 1D Ising model with a transverse magnetic field is given by
the Hamiltonian
\begin{eqnarray}
H=\lambda\sum_{i}^{N}\sigma_{i}^{x}\sigma_{i+1}^{x}+\sum_{i}^{N}\sigma_{i}^{z}.
\label{ham}
\end{eqnarray}
This model has a symmetry under a global rotation of $180^{\circ}$
over the $z$ axis ($\sigma^{x} \rightarrow -\sigma^{x}$) which
demands that $\langle \sigma^{x} \rangle = 0$. However as we
decrease the magnetic field, increasing $\lambda$, this symmetry
is spontaneously broken (in the thermodynamic limit) and we can
have a ferromagnetic phase with $\langle \sigma^{x} \rangle\neq0$.
This phase transition occurs at the critical point
$\lambda=\lambda_c=1$ where the gap vanishes and the correlation
length goes to infinity. This transition is named quantum phase
transition since it takes place at zero temperature and has many
of the characteristics of a second order thermodynamic phase
transition: phase transitions where the second derivative of the
free energy diverges or is not continuous. It is worth noting that
in the thermodynamic limit for $\lambda>1$ the ground state is
two-fold degenerated. These two states have opposite
magnetization. Here we will use the broken symmetric state for
$\lambda>1$ and not a superposition of the two degenerated states,
which is also a ground state but unstable. For a more detailed
discussion see Refs. \cite{nielsen,sachdev}.

Now, let us explain how we can evaluate $G(1)$ and $G(2,i_1)$ for
the one dimensional Ising model. We need, then, the reduced
density matrix of two spins, which is a $4\times 4$ matrix and can
be written as
\begin{eqnarray}
\rho_{ij}=\text{Tr}_{\overline{ij}}(\rho)=\frac{1}{4}\sum_{\alpha,\beta}
p_{ij}^{\alpha\beta}\sigma_{i}^{\alpha}\otimes\sigma_{j}^{\beta}.
\end{eqnarray}
The coefficients are given by
\begin{eqnarray}
p_{ij}^{\alpha\beta}=\text{Tr}\left(\sigma_{i}^{\alpha}\sigma_{j}^{\beta}
\rho_{ij}\right)=
\langle\sigma_{i}^{\alpha}\sigma_{j}^{\beta}\rangle, \label{pij}
\end{eqnarray}
and, as usual, $Tr_{\overline{ij}}$ is the partial trace over all
degrees of freedom except the spins at sites $i$ and $j$,
$\sigma_{i}^{\alpha}$ is the Pauli matrix acting on the site $i$,
$\alpha,\beta=0,x,y,z$ where $\sigma^{0}$ is the identity matrix,
and the coefficients $p_{ij}^{\alpha\beta}$ are real.

Eq.~(\ref{pij}) shows that all we need are the two-point spin
correlation functions which, in principle, are at most $16$. This
number can be reduced using the symmetries of the Hamiltonian
(\ref{ham}). The translational symmetry implies that $\rho_{ij}$
depends only on the distance $|i-j|=n$ between the spins so that
we have $p_{ij}^{\alpha\beta}=p_{n}^{\alpha\beta}$ and
$p_{n}^{\alpha\beta}=p_{n}^{\beta\alpha}$. All
these symmetries imply that the only non-zero correlation
functions are: $p_{n}^{\alpha\alpha}$, $p^{0x}=p^{x0}=p^{x}$,
$p^{0z}=p^{z0}=p^{z}$, and $p_{n}^{xz}=p_{n}^{zx}$.

First, let us show the diagonal correlation functions and the
magnetizations, which were already calculated in Ref. \cite{Ising
original}. For periodic boundary conditions and an infinite chain
we have:
\begin{eqnarray}
\langle\sigma_{i}^{x}\sigma_{i+n}^{x}\rangle= \left|
\begin{array}{cccc}
g(-1)  & g(-2)  & \cdots & g(-n)  \\
g(0)   & g(-1)  & \cdots & g(-n+1)\\
\vdots & \vdots & \ddots & \vdots \\
g(n-2) & g(n-3) & \cdots & g(-1)
\end{array}
\right|,
\end{eqnarray}

\begin{eqnarray}
\langle\sigma_{i}^{y}\sigma_{i+n}^{y}\rangle= \left|
\begin{array}{cccc}
g(1)   & g(0)   & \cdots & g(-n+2)\\
g(2)   & g(1)   & \cdots & g(-n+3)\\
\vdots & \vdots & \ddots & \vdots\\
g(n) & g(n-1)   & \cdots & g(1)
\end{array}
\right|,
\end{eqnarray}

\begin{eqnarray}
\langle\sigma_{i}^{z}\sigma_{i+n}^{z}\rangle=\langle\sigma^{z}
\rangle^{2}-g\left(n\right)g\left(-n\right),
\end{eqnarray}

\begin{eqnarray}
\langle\sigma^{z}\rangle=g\left(0\right),
\end{eqnarray}

and

\begin{eqnarray}
\langle\sigma^{x}\rangle= \left\{
\begin{array}{cc}
0 \, ,                      & \lambda\leq1\\
(1-\lambda^{-2})^{1/8} \, , & \lambda>1
\end{array}
\right., \label{sigma-x}
\end{eqnarray}
with
\begin{eqnarray}
g\left(n\right)=l\left(n\right)+\lambda l\left(n+1\right),
\end{eqnarray}
and
\begin{eqnarray}
l\left(n\right)=\frac{1}{\pi}\int_{0}^{\pi}dk\frac{\cos\left(kn\right)}{1+
\lambda^{2}+2\lambda \cos(k)}.
\end{eqnarray}

We are now left with the evaluation of $p_{n}^{xz}=p_{n}^{zx}$.
This calculation was made in Ref. \cite{McCoy} where the authors
obtained the off-diagonal, time and temperature dependent, spin
correlation functions. In the paramagnetic phase ($\lambda \le 1$)
the ground state has the same symmetries of the Hamiltonian which
leads to $p^{xz}_{n}=0$. For the ferromagnetic phase ($\lambda >
1$) an explicit evaluation leaves us with an expression in terms
of intricate complex integrals which are not straightforward to
compute. For this reason we will use bounds for this off-diagonal
correlation function.

\begin{figure}[ht]
\includegraphics[width=3in]{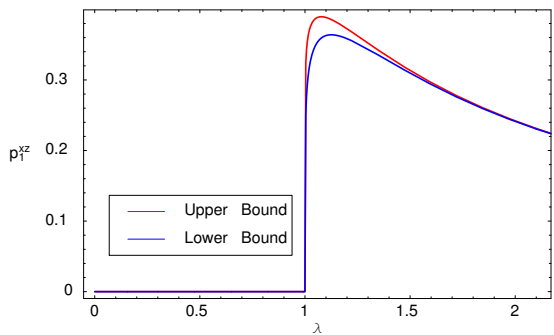}
\caption{\label{bounds_correl_xz} (Color online) Bounds for
$p^{xz}_{n}$ obtained by imposing the positivity of the eigenvalues 
of the reduced density operator $\rho_{ij}$.}
\end{figure}

We can obtain an upper and lower bound for this correlation
function by imposing the positivity of the eigenvalues of the
reduced density operator $\rho_{ij}$. For the Ising model these
bounds result to be very tight as we can see in Fig.
\ref{bounds_correl_xz}, and depend on $n$. In Ref.
\cite{nossopaper} some of the results here discussed were presented 
using zero as a lower bound. It is worth mentioning that since both
$G(1)$ and $G(2,i_1)$ are decreasing functions of the square of the 
correlation functions, a lower (upper) bound for the latter implies 
an upper (lower) bound for the former.

\begin{figure}[ht]
\includegraphics[width=3in]{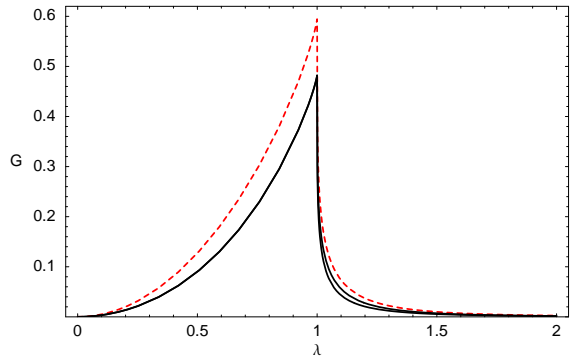}
\caption{\label{figuraG1_G21s} (Color online) G(1) (red/dashed
line) and the bounds for G(2,1) (black/solid lines). Note that they
are maximum at the critical point.}
\end{figure}

Since we have all the correlation functions at hand we proceed
with the calculations of $G(1)$ and $G(2,i_1)$. Remembering that
for the Ising model $p^y=0$ Eq.~(\ref{translation-g1}) can be
written as
\begin{eqnarray}
G(1)=1-(p^x)^2-(p^z)^2. \label{ising-g1}
\end{eqnarray}
As we have already shown $G(1)$ is the mean linear entropy of one
spin which, due to translational symmetry, is equal to the linear
entropy of any spin of the chain. A similar related analysis was
done by Osborne and Nielsen \cite{nielsen} for the  von Neumann
entropy instead of the linear entropy. As well as $G(1)$, see Fig.
\ref{figuraG1_G21s}, the von Neumann entropy  is maximal at the
critical point \cite{nielsen}. At that time Osborne and Nielsen
did not give much importance to this result since they suspected
that the von Neumann entropy of one spin with the rest of the
chain does not measure genuine MES. However, for a translational
symmetric state it is a reasonable good indication of genuine ME as we
have shown in previous sections. (We have explicitly studied
the linear entropy but the same results apply to the von Neumann
entropy. We have adopted the former mainly due to its
simplicity and relation to the Meyer and Wallach global
entanglement \cite{meyer}).

\begin{figure}[ht]
\includegraphics[width=3in]{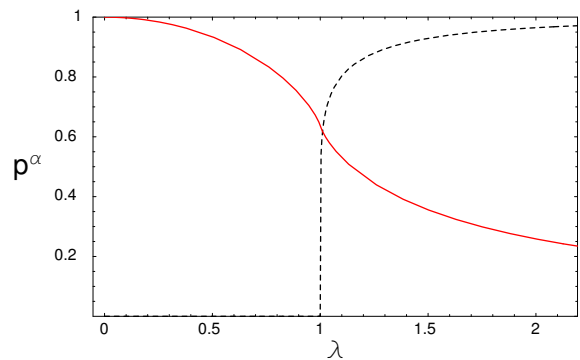}
\caption{\label{magnetizacoes_ising} (Color online) Magnetizations
$p^x$ $=$ $\langle \sigma^x \rangle$ (bla\-ck/da\-shed line) and
$p^z$ $=$ $\langle \sigma^z \rangle$ (red/solid line) as a function of
$\lambda$.}
\end{figure}

Analyzing Eq.~(\ref{ising-g1}) we can understand why $G(1)$ is
maximal at the critical point ($\lambda=1$). As we explain in what
follows, it is $\langle \sigma^x \rangle$ the main responsible for
this behavior of $G(1)$. For $\lambda \leq 1$ we have $\langle \sigma^x
\rangle =0$. After the critical point, however, $\langle \sigma^x
\rangle \neq 0$. Moreover, for $\lambda > 1$ Eq.~(\ref{sigma-x}) tells us 
that $\langle \sigma^x \rangle$ is a monotonic increasing
function of $\lambda$ and that $\langle \sigma^x \rangle
\rightarrow 1$ as $\lambda \rightarrow \infty$.  Therefore, since
$\langle \sigma^z\rangle$ is negligible for large values of
$\lambda$ and $\langle \sigma^x \rangle \approx 1$ (See Fig.
\ref{magnetizacoes_ising}) we must have $G(1)$ approaching zero
after the critical point.

\begin{figure}[ht]
\includegraphics[width=3in]{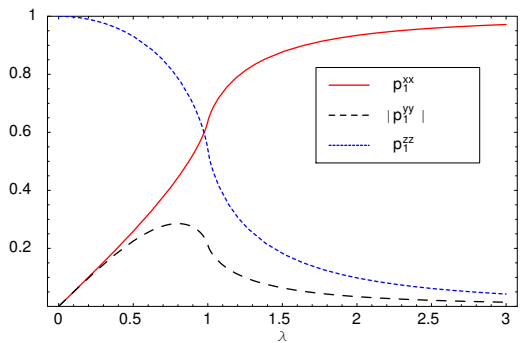}
\caption{\label{correlacoes_ising} (Color online) Two point
correlation functions: $p_1^{xx}$ (red/solid), $-p_1^{yy}$
(black/long-dashed), and $p_1^{zz}$ (blue/short-dashed).}
\end{figure}

We now analyze $G(2,i_1)$. Using the Ising model
symmetries Eq.~(\ref{translation-g2}) reads,
\begin{eqnarray}
G(2,n)=1-\frac{1}{3} \left[ 2(p^x)^2 + 2(p^z)^2 + 2(p^{xz}_{n})^2
+
\right. \nonumber \\
\left. (p^{xx}_{n})^2 + (p^{yy}_{n})^2 + (p^{zz}_{n})^2 \right].
\label{ising-g2}
\end{eqnarray}
With Eq.~(\ref{ising-g2}) we can evaluate $G(2,n)$ for any value
of $n$. In Fig. \ref{figuraG1_G21s} we have plotted $G(1)$ and the
bonds for $G(2,1)$. We can see that both $G(1)$ and $G(2,1)$ are
maximum at the critical point $\lambda=1$. Notice that the bounds
are very tight and can barely be distinguished just in a small
region for $\lambda\gtrsim 1$. Furthermore, $G(2,1)$ is always
smaller than $G(1)$, contrary to what was obtained using zero as
a lower bound \cite{nossopaper}.
As well as in the case of $G(1)$ we can see that the reason for
$G(2,1)$ being maximal at the critical point is due to the
behavior of $\langle\sigma^{x}\rangle$ since it is the only
function in Eq.~(\ref{ising-g2}) that does not change smoothly as
we cross the critical point (see Fig. \ref{correlacoes_ising} for
the other correlation functions).

\begin{figure}[ht]
\includegraphics[width=3in]{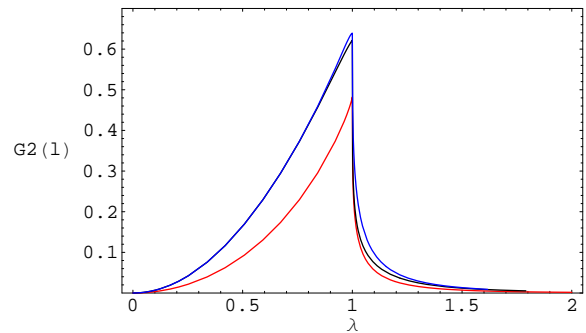}
\caption{\label{comparacao_G2sUB} (Color online) $G(2,n)$ for
$n=1,7$, and $15$. From bottom to top $n=1,7$, and $15$}
\end{figure}

\begin{figure}[ht]
\includegraphics[width=3in]{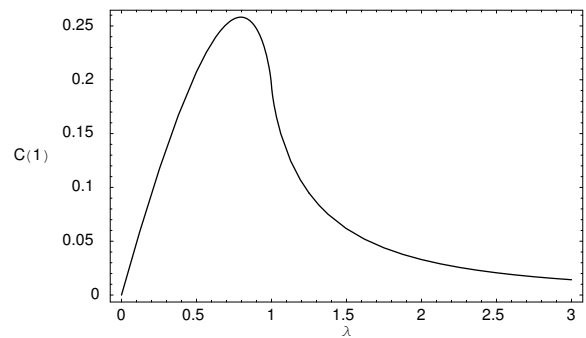}
\caption{\label{concorrencia} Concurrence for nearest neighbors.}
\end{figure}

We have also plotted $G(2,n)$ for $n=1$, $7$, and $15$ (Fig.
\ref{comparacao_G2sUB}). We can observe that all of them are
maximum at the critical point and increase as a function of $n$
(In Fig. \ref{comparacao_G2sUB} we have plotted only the upper
bounds since the lower bounds produce very similar curves). We
also note that $G(2,7)$ is very near $G(2,15)$ showing that
$G(2,n)$ rapidly saturates to a fixed value. At the critical point
we have $\lim_{n\rightarrow\infty}G(2,n)=0.675$. This behavior for
$G(2,n)$ points in the direction of the existence of multipartite
entanglement at the critical point since any two spins are
entangled with the rest of the chain and this entanglement
increases with the distance between them. It is also interesting
to confront this result with the fact that two spins that are
separated by two or more sites are not entangled since their
concurrences are zero \cite{Nature}.
The behavior of the concurrence ($C(n)$) can also be understood if
we note that it can be expressed in terms of the one and two point
correlation functions. While for the non-symmetric (ferromagnetic)
state  the analytical expression for the concurrence is cumbersome
for the symmetric one it is very simple. Fortunately, for the
Ising model it was show that the concurrence does not change upon
symmetry break \cite{nome_dificil1,nome_dificil2} and it turns out to
be
\begin{eqnarray}
C(n)=\frac{1}{2} \left(-1-p_n^{yy}+p_n^{xx}+p_n^{zz} \right).
\end{eqnarray}
From this expression we can see that the concurrence (Fig.
\ref{concorrencia}) does not depend on either the off-diagonal
correlation function $p_n^{xz}$ or on the one point correlation
functions (magnetizations). This is an interesting feature and
helps us to understand why the concurrence is not maximum at the
critical point.

\section{Conclusion}
\label{conclusion}

A $N$-partite quantum system may be entangled in many distinct
ways. To characterize and to define a good measure of entanglement
for those systems is a hard problem. The only simple alternative,
valid whenever the joint $N$-system state is pure, is to split the
system into two partitions and compute the entanglement in that
way. This bipartition could be constructed in many different forms
and thus give distinct amount of entanglement. One possible
approach is to divide the system into two blocks of $L$ and $N-L$
subsystems and to compute the block entanglement
\cite{latorre,latorre2} between the two blocks. However one could
think of a situation where all of the subsystems in the block $L$
are entangled with each other, as well as the subsystems of block
$N-L$, but without any entanglement between the two blocks. For
this situation the block entanglement would quantify a zero amount
of entanglement, which is clearly not true. A valid bipartition
approach, which would be able to quantify the entanglement in such
a situation, is to compute the entanglement for all kinds of
bipartition and then to average these to give the total amount of
entanglement in the system.

In this article we have formalized an operational multipartite
entanglement measure, {\it the generalized global entanglement}
($E_{G}^{(n)}$), firstly introduced in Ref. \cite{nossopaper}. For
$n=1$, $E_{G}^{(n)}$ recovers the Meyer and Wallach global
entanglement measure \cite{meyer}. However for $n>1$ $E_{G}^{(n)}$
together with the auxiliary function $G(n, i_1, i_2, \ldots,
i_{n-1})$ quantify  entanglement in the many distinct forms it is
distributed in a multipartite system. We have shown that for some
multipartite systems the original global entanglement is not able
to properly classify and identify multipartite entanglement in a
unequivocally way, whereas higher classes ($n>1$) of $E_G^{(n)}$
are. A genuine $k$-partite entangled state is the one that cannot
be written as a product $|\phi\rangle_l \otimes
|\psi\rangle_{(k-l)}$ of state vectors for any $l<k$, meaning that
there is no other reduced pure state out of the joint $k$-systems
state. To completely quantify and classify the multipartite
entanglement in this kind of state one would have to compute all
the $E_{G}^{(n)}$ classes up to $n=k-1$. However we have observed
that lower classes of $E_{G}^{(n)}$, such as $E_{G}^{(1)}$ and
$E_{G}^{(2)}$, are sufficient to detect multipartite entanglement.
The computation of higher orders of $E_{G}^{(n)}$ and of the
auxiliary functions $G(n, i_1, i_2, \ldots, i_{n-1})$ is
necessarily required only to distinguish and classify the ways the
system is entangled. Although  the calculation of all those higher
orders may be operationally laborious it is straightforward to
perform for finite $N$ systems. Thus we have demonstrated for a
variety of genuine multipartite entangled qubit states
\cite{osterloh,chua} that $E^{(2)}_G$ and $G(2,i_1)$ are able to
properly identify and distinguish them whereas $E_G^{(1)}$ fails to do so.
Inspired by the common characteristic presented by all $G(2,i_1)$
for those paradigmatic states we then discussed an operational
definition of a genuine multipartite entangled state
\cite{osterloh,chua}.

Finite multipartite systems are interesting
for fundamental discussions on the definition of multipartite
entanglement. Infinite systems on the other hand are interesting
since multipartite entanglement may be relevant to improve our
knowledge of quantum phase transition processes occurring in the
thermodynamical limit. We have demonstrated that for the 1D Ising
model in a transverse magnetic field both $E_G^{(2)}$ and
$G(2,i_1)$ are maximal at the quantum critical point,
suggesting thus a favorable picture for the occurrence of a
genuine multipartite entangled state. Moreover, the behavior of
$G(2,i_1)$ and thus $E_G^{(2)}$ can be easily understood as
contributions of the one and two-point correlation functions 
giving us a physical picture for the behavior of the 
multipartite entanglement during the phase transition process.

In conclusion the generalized global entanglement we presented has
the following important features: (1) It is operationally easy to
be computed, avoiding any minimization process over a set of
quantum states; (2) It has a clear physical meaning, being for
each class $E_G^{(n)}$ the averaged $n$-partition purity; (3) It
is able to order distinct kinds of multipartite entangled states
whereas other common measures fail to do so; (4) It is able to
detect second order quantum phase transitions, being maximal at the
critical point. (5) Finally, for two-level systems it is given in terms of 
correlation functions, and thus easily computed for a variety of
available models. We hope that this measure may contribute for
both the understanding of entanglement in multipartite systems
and for the understanding of the relevance of entanglement in
 quantum phase transitions.

\begin{acknowledgments}
GR and TRO acknowledge financial support from Funda\c{c}\~ao de
Amparo \`a Pesquisa do Estado de S\~ao Paulo (FAPESP) and MCO
acknowledges partial support from FAPESP, FAEPEX-UNICAMP and from
Conselho Nacional de Desenvolvimento Cient\'\i fico e
Tecnol\'ogico (CNPq).
\end{acknowledgments}


\begin{thebibliography}{50}
\bibitem{schroedinger} E. Schr\"{o}dinger, Proc. Camb. Phil. Soc. \textbf{31},
 555 (1935).
\bibitem{epr} A. Einstein, B. Podolsky, and N. Rosen, Phys. Rev. \textbf{47},
 777 (1935).
\bibitem{bell} J. S. Bell, Physics \textbf{1}, 195 (1964).
\bibitem{livrodonielsen} M. A. Nielsen and I. L. Chuang, Quantum
Computation and Quantum Information (Cambridge University Press,
Cambridge, 2000).
\bibitem{wootters} W. K. Wootters, Phys. Rev. Lett. \textbf{80},
2245 (1998).
\bibitem{Schmidt} S. L. Braunstein and P. van Loock, Rev. Mod. Phys.
\textbf{77}, 513 (2005).
\bibitem{cirac} W. D\"{u}r, G. Vidal, and J. I. Cirac, Phys. Rev. A
\textbf{62}, 062314 (2000).
\bibitem{verschelde} F. Verstraete, J. Dehaene, B. De Moor, and H. Verschelde,
Phys. Rev. A \textbf{65}, 052112 (2002).
\bibitem{footnote1} Richer classifications with more parameters can
be pursued. For instance, we can have a multipartite state where
four subsystems are entangled, another five are entangled, and the
remaining subsystems are separable. We would need now two
parameters to classify this state.
\bibitem{nossopaper} T. R. de Oliveira, G. Rigolin, and M. C. de Oliveira, 
Phys. Rev.A \textbf{73}, 010305(R) (2006).
\bibitem{meyer} D. A. Meyer and N. R. Wallach, J. Math. Phys. \textbf{43},
4273 (2002).
\bibitem{viola} H. Barnum, E. Knill, G. Ortiz, R. Somma, and L. Viola,
Phys. Rev.Lett. \textbf{92}, 107902 (2004).
\bibitem{somma} R. Somma, G. Ortiz, H. Barnum, E. Knill, and L. Viola,
Phys. Rev. A \textbf{70}, 042311 (2004).
\bibitem{brennen} G. K. Brennen, Quantum Inf. Comp. \textbf{3}, 619 (2003).
\bibitem{indianos} A. Lakshminarayan and V. Subrahmanyam,
Phys. Rev. A 71, 062334 (2005).
\bibitem{latorre} J. I. Latorre, E. Rico, and G. Vidal, Quantum Inf. Comp.
\textbf{4}, 48 (2004) and references therein.
\bibitem{latorre2} G. Vidal, J. I. Latorre, E. Rico, and
A. Kitaev, Phys. Rev. Lett. \textbf{90}, 227902 (2003).
\bibitem{CommentScott} We remark that A. J. Scott in Ref.
\cite{scott} has previously arrived to a simmilar generalization
of the Meyer-Wallach global entanglement from a somewhat different
approach. In our form however, due to the definition in terms of
the auxiliary function $G(n, i_1, i_2, \ldots, i_{n-1})$, the many
facets of the multipartite entanglement are clearly represented
through the  classes $n$ of $E_G^{(n)}(\rho)$.
\bibitem{scott} A. J.
Scott, Phys. Rev. A \textbf{69}, 052330 (2004).
\bibitem{nome_dificil1} O. F. Sylju{\aa}sen, Phys. Rev. A \textbf{68},
060301(R) (2003).
\bibitem{nome_dificil2} O. F. Sylju{\aa}sen, eprint quant-ph/0312101.
\bibitem{ghz}D. M. Greenberger, M. A. Horne, A. Shimony, and A. Zeilinger, Am.
J. Phys. \textbf{58}, 1131 (1990).
\bibitem{rigolintele}G. Rigolin, Phys. Rev. A \textbf{71}, 032303 (2005).
\bibitem{osterloh} A. Osterloh and J. Siewert, Phys. Rev. A \textbf{72},
012337 (2005).
\bibitem{chua} Y. Yeo and W. K. Chua, Phys. Rev. Lett. \textbf{96}, 060502
(2006).
\bibitem{coffman}V. Coffman, J. Kundu, and W. K. Wootters, Phys. Rev. A
\textbf{61}, 052306 (2000).
\bibitem{guo}  C. P. Yang and G. C. Guo, Chin. Phys. Lett. \textbf{17}, 162
(2000).
\bibitem{negativity} G. Vidal and R. F. Werner, Phys. Rev. A \textbf{65},
032314 (2002).
\bibitem{footnote2} Another way of interpreting Definition \ref{def1} is to
consider it as a sufficient condition for a state to be a genuine
MES.
\bibitem{nielsen}T. J. Osborne and M. A. Nielsen, Phys. Rev. A \textbf{66},
032110 (2002).
\bibitem{Nature}A. Osterloh, L. Amico, G. Falci and R. Fazio, Nature 416, 608
(2002).
\bibitem{sachdev} S. Sachdev, \textit{Quantum Phase Transitions} (Cambridge 
University Press, Cambridge, 1999).
\bibitem{Ising original}P. Pfeuty, Ann. Physics (New York) \textbf{57},
79 (1970).
\bibitem{McCoy}J. D. Johnson and B. M. McCoy, Phys. Rev. A \textbf{4}, 2314
(1971).
\bibitem{tognetti} T. Roscilde, P. Verrucchi, A. Fubini, S. Haas, and 
V. Tognetti, Phys. Rev. Lett. \textbf{94}, 147208 (2005).
\bibitem{verstraete} F. Verstraete, M. A. Martin-Delgado, and J. I. Cirac,
Phys. Rev. Lett. \textbf{92}, 087201 (2004).
\bibitem{venuti} L. Campos Venuti, C. Degli Esposti Boschi, M. Ron\-ca\-glia, 
and A. Scaramucci, Phys. Rev. A \textbf{73}, 010303(R) (2006).
\end{thebibliography}
\end{document}